# Development of an 8 channel sEMG wireless device based on ADS1299 with Virtual Instrumentation

**Marcelo Bissi Pires [1], José Jair Alves Mendes Junior [2] and Sergio Luiz Stevan Jr[2,3]***

[1]**Abstract:** In this paper, a different approach on the use of the ADS1299 (an analog front-end with features for electroencephalogram and electrocardiography signal acquisition) is considered, proposing the development of a surface electromyography (sEMG) device. The main features of the device include simultaneous recordings of eight muscular channels, wireless transmission and virtual instrumentation with the use of LabVIEW[TM] software. The proposed sEMG device contains a specifically designed protocol to accommodate data transmission by reducing the data size while still delivering adequate resolution (34.33 μV), amplitude range (±17.57 mV) and sampling rate (1000 Hz) for sEMG signals. For the validation methods, a generated sine wave and a known sEMG data were evaluated. Moreover, the muscular recordings for all the eight channels of a human arm were successful and the results expose the isolated contractions of the triceps and the biceps with their amplitude range and frequency spectrum.

**Keywords:** Electromyography; ADS1299; LabVIEW; sEMG; wireless.

## 1. Introduction

In the last decades, the improvements of the technology on devices used in health care have been very important to guarantee a better quality of life to people with health problems [1]. In this subject, the surface electromyography (sEMG) is a non-invasive technique used to capture muscle signals from the surface of the skin of an individual. This typical biomedical signal is generated by the contraction and expansion of muscles [2] and has been proved useful in a handful of medical and electronic research areas.

The sEMG technique may aid in the diagnostic of muscle disorders, such as low back pain [3], fibromyalgia [4] and tremors [5]. Furthermore, it can also be useful in rehabilitation by assessing some muscle parameters such as fatigue [6], motion [7], and by providing prosthetic devices [8]. There are also scientific papers that go beyond the medical area such as a sign language recognition system [9], a human amplification device with the help of an exoskeleton [10], a facial expression monitoring system [11], and an eyeglasses prescription method by means of sEMG [12].

Regardless of the area of study or the final objectives to be achieved when analyzing the sEMG signal, a reliable acquisition system is of great importance to deliver consistent and accurate data. The acquisition system is responsible for delivering the sEMG signal to an environment where it can be analyzed and processed digitally as needed. In most cases, the main components involved to capture a posture or a muscular contraction are surface electrodes, signal amplifiers, filters and an analog-to-digital (A/D) converter [13].

Taking into account that such components can be used in many other electronic devices, it is predictable that they have a large variety of models and values. As a result of that, sEMG acquisition systems may have distinct aspects from one another such as the number of channels, sample rate, filter bandwidth, A/D resolution, gain and many others.

In this scenario, this paper brings forward the development of a multichannel sEMG wireless device, which relies mainly on the use of the Texas Instruments analog front-end ADS1299 and the National Instruments programming environment named LabVIEW$^{TM}$. This device relies on a particular communication protocol to maximize data transmission, providing adequate sampling rate, full-scale range, and resolution for sEMG signals. In addition, a small review of the many acquisition devices that are also based on the ADS1299 is presented, exposing their main features and purpose of the application. The main differences from the proposed sEMG device and the review rely essentially on the number of channels used, application, signal validation, portability, the presence of a user interface, and overall technical details when describing the hardware and communication protocol used.

This paper is organized as follows. The related work is reviewed in section 2. The proposed sEMG device is introduced and described thoroughly in section 3, with the detailing of hardware, the communication protocol, the software programming, and the used methodology. In section 4, validation (using a known sine wave and recorded sEMG data) and recording (sEMG from arm muscles) results are exposed and discussed. Finally, in section 5, the overall performance of this work is asserted and additional details on further development are provided.

**2. Review of the acquisition devices based on the ADS1299**

The following literature review is based on an online search on recognized international journals, selecting articles from the year of 2013 until 2017. There are only a few numbers of articles that describe the acquisition system thoroughly and even though these would be the most desired in this review, many others that simply uses the ADS1299 for biomedical studies were also considered as long as they provided sufficient details about their device. The purpose of this review is to explore the articles that bring specific details on the hardware development used to make the acquisition devices, allowing a better overview on this subject, exposing whether the ADS1299 is commonly used for sEMG recordings.

In the research field of biomedical engineering, there are a few design concepts that guide the development of acquisition systems used in health care. The number of channels available is related to the classification accuracy. Devices that are associated with electromyography (EMG), electroencephalogram (EEG) or electrocardiography (ECG), generally have more than one channel available for recordings.

In addition, the quality of the biomedical signal depends on the filter capabilities to remove unwanted slow signal variations and noises. Moreover, the higher the resolution of the A/D converter, the better the device is to detect subtle changes in the signal. Likewise, the sampling rate must be high enough so the signal can be properly reproduced by satisfying the Nyquist sampling criterion.

Having in mind that electrodes used in sEMG, EEG, and ECG are in contact with the skin it is important that the device is safely isolated from the patient. In this matter, the use of low voltage batteries helps to increase the safety conditions while also granting a better mobility to the device. Similarly, the presence of wireless data transmission technologies such as Wi-Fi and Bluetooth grants portability and another layer of electrical isolation. Moreover, graphical user interfaces (GUI) are useful tools that provide easy accessibility and interaction with the device, delivering visual indicators of the biomedical signal (graphics) as well as other functions such as saving and processing data.

Consequently, a microcontroller or an FPGA (Field Programmable Gate Array) is important to control, process and transmit data as required. For that reason, low power consumption, high processing power, and intrinsic communication protocols are characteristics that benefit the acquisition device by providing reliable operation.

Therefore, the review shown in Table 1 includes most common aspects that are present in most of the devices reviewed. The authors and the year (A/Y) of the article are in the first column. In the second column, the purpose of the study is briefly asserted. On the following columns there are: the number of channels in differential mode (CH), the hardware filter bandwidth (FBW), the resolution of the A/D converter (RES), the sampling frequency (SR), the presence of a graphical user interface (GUI), battery supply (BS), the microcontroller used (μC) and if there is wireless capability present (WR). Whenever there are not enough details or the authors make no mention of a feature in the acquisition device, a dash is used in the table to represent such situation.

In some articles, the acquisition device is described in a particular way, however, when the recordings were performed, the mentioned details (aspects) are modified, using different values or methods to deliver the results. It was noted that different settings are generally used as a way to reduce the overall performance, such as a lower sampling rate than the maximum possible or a lower signal resolution. Although this decision to not use the full potential of the acquisition device is not clearly justified in some articles, it is believed to be because of other limitations that might appear later on in the signal processing chain.

For example, a lower sampling rate delivers fewer data to handle and consequently can be transferred over a communication system faster. However, in this example, the sEMG signal may or may not be compromised by this and thus it all comes down to a matter of trade-offs. With all that in mind, the characteristics presented in Table 1 take into consideration only the hardware settings used to obtain the results.

**Table 1.** Acquisition devices based on the ADS1299 front-end.

| A/Y | Purpose | CH | FBW | RES | SR | GUI | BS | μC | WR |
|---|---|---|---|---|---|---|---|---|---|
| [11] 2013 | sEMG | 4 | - | 24 bits | 1 kHz | LabVIEW | Yes | Arduino Uno | WiFi |
| [14] 2015 | Eye Movements | 2 | - | 24 bits | 250 Hz | - | No | Arduino | No |
| [15] 2014 | EEG | 8 | - | 24 bits | 250 Hz | C++ | - | XC6SLX100 | No |
| [16] 2013 | EEG | 4 | - | 24 bits | 250 Hz | LabVIEW | Yes | Arduino Pro Mini | Zigbee |
| [17] 2016 | EEG | 8 | 0-30 Hz | 24 bits | 250 Hz | - | - | STM32F103VET6 | WiFi |
| [18] 2017 | EEG | 8 | 0-20 Hz | - | - | OpenBCI | - | Arduino UNO | - |
| [19] 2013 | EEG | 8 | - | 24 bits | - | - | - | dsPIC33 | Bluetooth |
| [20] 2015 | sEMG | 4 | 5 Hz-∞ | - | 1 kHz | - | - | MSP430 | Bluetooth |
| [21] 2014 | EEG | 8 | - | 24 bits | 1 kHz | - | Yes | PIC32MX775512L | No |
| [22] 2015 | sEMG | 4 | 5-450 Hz | - | 2 kHz | - | No | - | No |
| [23] 2015 | EEG | 8 | - | 24 bits | 1 kHz | Matlab | Yes | AT32 Atmel AVR | Bluetooth |
| [24] 2015 | sEMG | 8 | - | 24 bits | 500 Hz | - | Yes | - | Bluetooth |

Analyzing the Table 1, it is clear that there are only a few numbers of articles that use the ADS1299 as an analog front-end specifically for capturing muscle signals and a fewer number describing the board or circuit in great detail. Moreover, these few articles are missing important aspects when describing the device, such as the resolution per bit (in volts), transmission baud rate or if there is a GUI or not.

Therefore, the proposed sEMG device is described in greater detail in this paper and brings forward an effective communication protocol designed to deliver higher sampling rate (1000 samples per second) and portability to record eight channels of muscular activity simultaneously. On a side note, signal validation (included in this work) is an important step to verify the consistency and accuracy of data acquisition devices which is not present in any of the articles reviewed.

## 3. Materials and Methods

The following subsections bring in great detail the main specifics of the sEMG device such as electronic components and diagrams, hardware architecture, communication protocol development, virtual instrumentation and others.

### *3.1. Proposed sEMG hardware development and aspects*

The wireless sEMG device developed in this study is mainly composed of: electrodes, filters, ADS1299 front-end, Bluetooth, microcontroller, power supply, a printed circuit board (PCB) and a GUI. Therefore, this section brings forward the details of these components, elucidating their purpose in the device and overall aspects.

Figure 1 depicts the arrangement of the main components and the data flow. The gray arrows represent data flowing in a wired communication, while the green arrow represents a wireless transmission. The red area indicates the hardware components that are powered by the batteries while the blue area represents the components that are mounted on the PCB. Regarding the sEMG recordings, initially, the electrodes transmit the biopotentials from the muscles to the filter, which then are converted to digital data by the front-end ADS1299 and a microcontroller. Lastly, by means of a wireless system (Bluetooth) the sEMG signal is transmitted to a personal computer with a properly designed GUI.

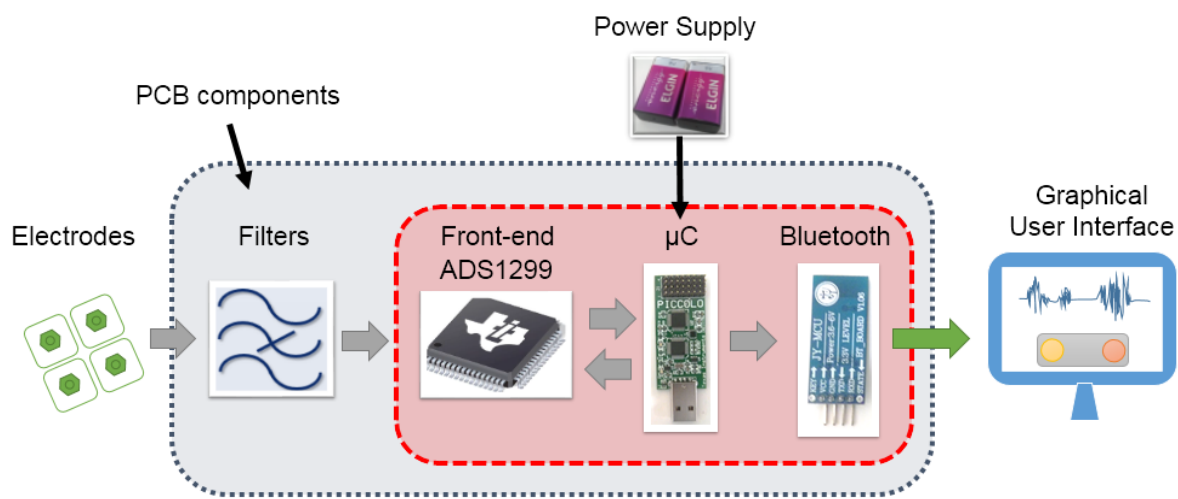

**Figure 1**. Proposed sEMG acquisition device.

*3.1.1. Electrodes*

In sEMG recordings, the electrodes used to measure muscular activity are generally of two types: dry or wet. Wet electrodes require an electrolyte gel to be applied on the skin surface to increase conductivity. On the other hand, dry electrodes usually have skin preparation such as shaving and cleaning prior to the recordings to reduce skin-electrode impedance [13]. In this work, generic patch-based dry electrodes were used.

*3.1.2. Filter Module*

Considering that electronic devices are subject to noises and interferences, the use of filters is essential to attenuate undesired frequencies, increasing the signal interpretation. Figure 2a shows a schematic of the filters developed in this study and Figure 2b presents the frequency spectrum of the filter (blue area indicates the passband and the red areas, the stopbands).

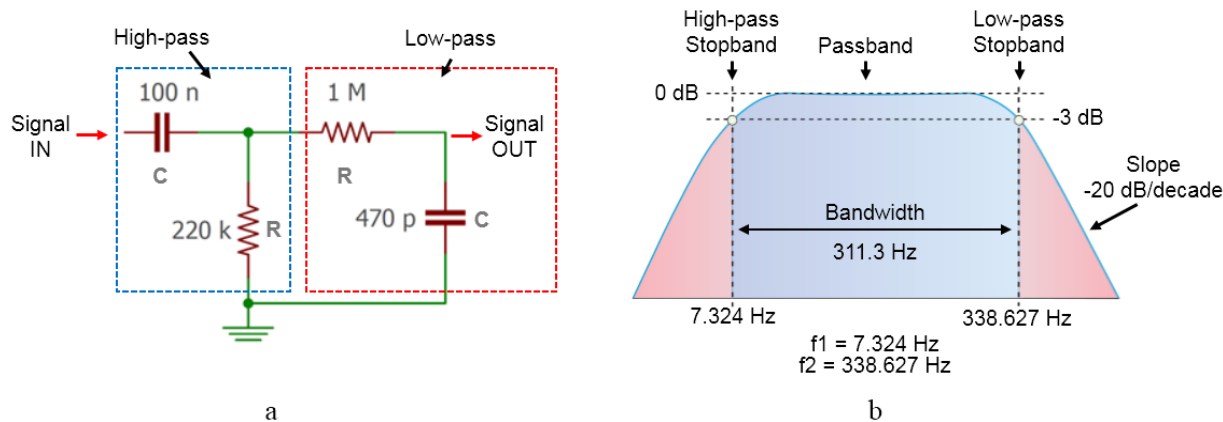

**Figure 2**. Band-pass filter module: (**a**) electronic schematic and (**b**) frequency spectrum.

A resistor-capacitor (RC) band-pass filter was designed combining a high-pass and a low-pass filter and arranged after the electrodes to attenuate noises that occur in the skin-wire-electrode interface. Considering that the sEMG frequency spectrum ranges from 0 to 400 Hz [25], the low-pass stage of the filter has a cutoff frequency of 338.627 Hz. However, the aforementioned low-frequency noises, such as motion artifact and electrochemical voltage, [25] may contaminate the sEMG signal. Therefore, a high-pass filter with a cutoff frequency of 7.234 Hz was designed to attenuate such interferences.

*3.1.3. Front-end ADS1299*

The analog front-end ADS1299 is a low-noise delta-sigma A/D converter with an intrinsic programmable gain amplifier (PGA). This particular integrated circuit (IC) available since 2012 from Texas Instruments delivers all analog components necessary for EEG and ECG applications [26]. There is no specific mention about sEMG applications in the datasheet, however, considering the nature of biopotentials, this study still benefits from the use of the ADS1299 to produce sEMG recordings.

The main features of this IC are the following: up to 8 recording channels, unipolar or differential input mode, 24-bits resolution, gain of up to 24 (PGA), unipolar or bipolar supply, internal A/D reference of 4.5 Volts and data rate from 250 to 16000 samples per second (SPS). For this particular work, the ADS1299 was configured as described in Table 2.

**Table 2.** Acquisition devices based on the ADS1299 front-end.

| Feature | Configuration |
|---|---|
| Channels | 8 |
| Input mode | Differential |
| PGA | 1 |
| Supply mode | Bipolar ± 2.5 V |
| Internal A/D reference | On |
| Data rate | 1000 SPS |

*3.1.4. Front-end ADS1299*

Acquisition devices that are physically connected to a computer or a display restrict the movement range of the subjects. Moreover, unavoidable displacement disturbances in the wires contribute to interferences, lowering the signal quality. Therefore, to deliver a better portability and more practical recordings, wireless communication is used.

Amongst all possible wireless communications and modules, the JY-MCU HC-06 is used in this study for being a low-cost Bluetooth 2.1 module [27]. We could have used a Bluetooth module 4, with lower power consumption, however, the JY-MCU HC-08 low-cost module presents many reported problems with Android devices, which we consider an important limitation for use in displacement and remote acquisition. An important point is that the speed of Bluetooth 2.1 supplied our need for signal sampling with frequencies less than 1kHz (double of cut-off frequency). Furthermore, most recent portable electronic devices (such as laptops and smartphones) are equipped with a Bluetooth receiver, which is an additional benefit to the proposed sEMG device in terms of compatibility. This module transmits the digital sEMG data to a computer wirelessly. According to the datasheet, the maximum baud rate in this situation is 115200 bits per second (bps) and consequently, this is the baud rate used in this work.

*3.1.5. Microcontroller*

The central controller of this device is the C2000 Piccolo ControlSTICK TMS320F28069 microcontroller from Texas Instruments [28]. The C2000 is responsible for configuring the ADS1299 while also managing the overall data flow. The characteristics of the TMS320F28069 are adequate for this project because it includes a USB (Universal Serial Bus) interface, 90 MHz clock, 32-bits Central Processing Unit, 256 KB of flash memory, Serial Peripheral Interface (used by us to communicate with the ADS1299) and Serial Communication Interface (resource used by us to communicate with the Bluetooth) modules and C/C++ programming language. Although this microcontroller was chosen, many others could be used, as long as they have the aforementioned features.

*3.1.6. Power supply*

In a biomedical device, the safety of the subject is a priority. The power circuit designed aimed not only to deliver all the necessary voltages to the components but also to isolate the device from external high power sources such as the electrical grid. Therefore, two common 9 Volts batteries and linear voltage regulators were used to supply the sEMG device. The purpose of having two 9 Volts batteries is to provide different supply rails (positive and negative) without the necessity of switching regulators (using only one 9 Volts battery). The use of two batteries increases the current capacity, but due to a small different current drawing from each trail, voltage source stability is affected when the batteries are close to reaching their depletion. As a future alternative, an integrated circuit such as the MAX1044 / ICL7660, or an auxiliary circuit based the LM555 CI could be used for input voltage mirroring and thus provide the necessary negative voltage for the electronic circuit. However, these options were not chosen as an option to the greater autonomy of the circuit.

A schematic of the power supply circuit is depicted at the left bottom of the Figure 3a. The LM7805, LM317, and LM337 regulators are responsible for proving the voltages required for the ADS1299 (+2.5V and -2.5V) and for the C2000 and the JY-MCU (+5V).

*3.1.7. Circuit diagram*

The electronic layout of the PCB in shown in Figure 3a. Essentially, all 16 electrodes are connected to their own RC filter. Each pair of filter outputs (negative and positive) are connected to the ADS1299, creating all the eight sEMG channels. The microcontroller (C2000) is connected to the appropriate ports of the ADS1299 as well as to the JY-MCU, arranging sEMG data for transmitting. The C2000 can be powered either by the batteries or directly by a USB port. Therefore, a jumper must be removed from the USB_JP pins to toggle off the power coming to the C2000 when powering the microcontroller from a USB port. This way, a different firmware can be uploaded if necessary. The Bluetooth and C2000 peripheral modules are shown at the top of the Figure 3b. Similarly, the top and bottom layer of the manufactured PCB (8.4 cm wide and 6.6 cm tall) is presented at the bottom of the Figure 3b.

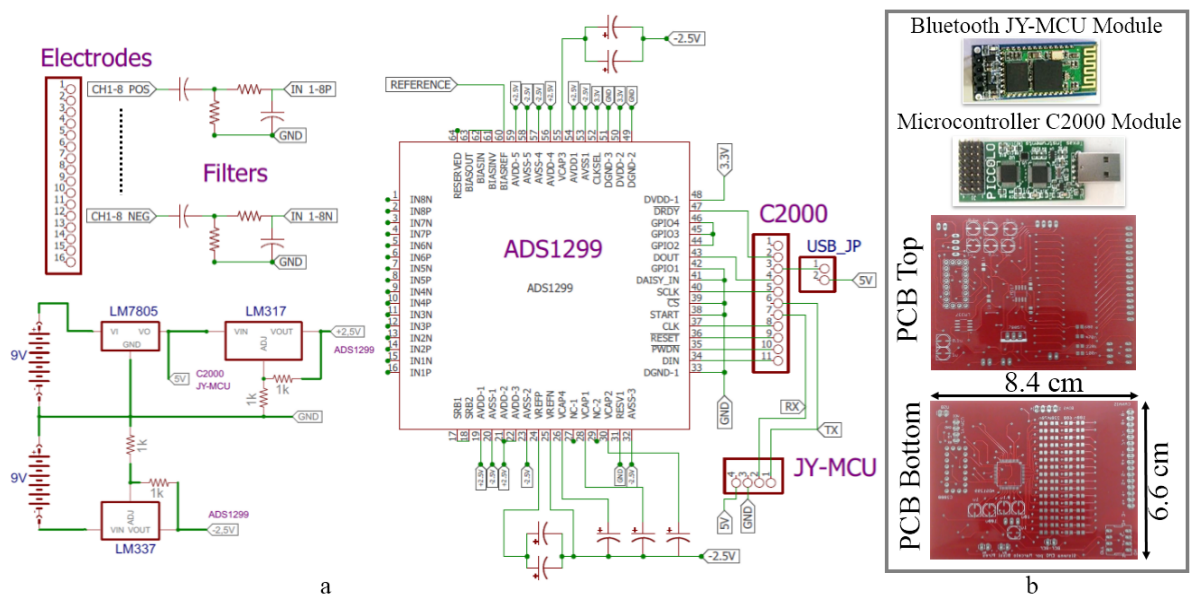

**Figure 3**. Printed circuit board: (**a**) electronic schematic and (**b**) peripherals modules (Bluetooth JY-MCU and Microcontroller C2000) and PCB top and bottom layer.

## 3.2. Communication Protocol

The performance of a wireless communication protocol for sEMG devices is usually related to three important aspects: the resolution of the A/D converter, the maximum baud rate of the Bluetooth and the number of recording channels.

In this scenario, the data frame for asynchronous serial communication is composed of 10 bits in total (1 start bit followed by 8 data bits and 1 stop bit). Therefore, the data of a single sEMG channel contains 24 data bits, 3 start bits and 3 stop bits for a total 30 bits to be transmitted. As a consequence, in the proposed sEMG device, the amount of data necessary to transmit all the eight channels is 240 bits. At the maximum rate of the Bluetooth module (115200 bps), the throughput (TP) for this situation is

$$\text{TP} = \frac{\text{Baud rate}}{\text{Frame length}} = \frac{115200 \text{ bps}}{240 \text{ bits}} = 480 \text{ Hz}. \quad (1)$$

As mentioned in Table 2, the sample rate of the ADS1299 was set to 1000 SPS. However, the wireless transmission rate in (1) severely restricts the sample rate of the sEMG device, and as a consequence, the Nyquist criterion is not satisfied.

As an alternative, the sample rate can be increased by reducing the length of the message which, as a consequence, lowers the resolution and the full-scale range of the sEMG signal. The amplitude of the sEMG signal usually ranges from microvolts [30, 31] up to ± 10 mV [13].

In this scenario, a wireless communication protocol was designed specifically for this device, reducing the amount of data necessary for transmission, assuring a sampling rate of 1000 SPS while covering the amplitude range of sEMG signals.

The resolution per bit for the ADS1299 is 0.53644 μV considering the internal voltage reference of 4.5 V and the 24 bits resolution (the most significant bit - MSB - is only used to determinate the sign). However, the designed protocol reduces the amount of data transmitted per channel, from 24 bits to 10 bits (1 sign bit and 9 data bits). Additionally, a character (composed of 8 bits) is transmitted after all the eight channels to indicate the end of the message, resulting in a throughput of approximately 1047.27 Hz. In this particular case, the wireless communication is faster than the sampling rate of the ADS1299 satisfying the Nyquist criterion in the whole processing chain.

The reduction in the data size is based on a proper selection of the bits present in the 24-bit A/D conversion. There are many possible data windows, and one of them is presented in Figure 4. In this arbitrary data of a single sEMG channel (24 bits), the green box represents the sign bit (0 for negative and 1 for positive) and the yellow box represents the magnitude of the selected window. This particular window, designated for this sEMG device, is capable of delivering adequate resolution and amplitude range for sEMG signals based on a trade-off between larger amplitudes (moving the window to the left) and better resolution (moving the window to the right).

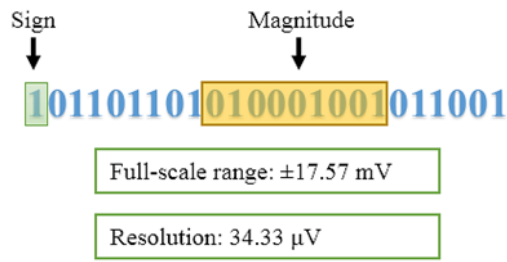

**Figure 4**. Data window selection for a 24 bits sEMG channel.

In order to encrypt data for the protocol, binary operations are used in the firmware to select the proper bits before transmitting them. Figure 5a depicts this scenario using an arbitrary data as an example. This protocol relies on two data structures: the least significant package (LSP) and the most significant package (MSP).

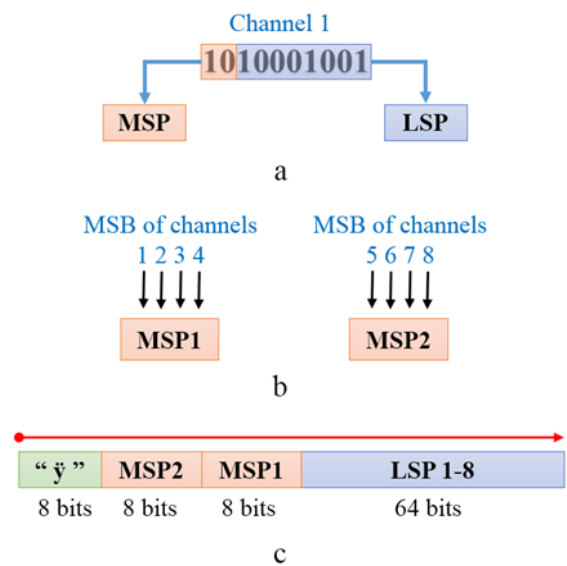

**Figure 5**. Encryption of data, with (**a**) data structures, (**b**) formation of each MSP, and (**c**) length of one sample.

The LSP (blue box) is unique for each channel and contains the eight least significant bits of the selected data window. In contrast, as exposed in Figure 5b, each MSP (pink box) contains information from 4 different channels at once (two most significant bits from each window), increasing efficiency. Therefore, for a single sample of all channels, eight LSPs are transmitted first, followed by two MSPs and the end-of-message character "ÿ", respectively, as shown in Figure 5c. As a result, there are 64 bits in all eight LSPs, 16 bits in two MSPs and 8 bits in the character for a grand total of 88 data bits, or 110 bits including overheads, in the whole message.

For this particular message length, considering an 115200 bps transmitting speed, the throughput (TP) is approximately

$$\text{TP} = \frac{\text{Baud rate}}{\text{Frame length}} = \frac{115200 \text{ bps}}{110 \text{ bits}} = 1047.27 \text{ Hz}. \quad (2)$$

The throughput calculated in (2) surpasses the ADS1299 sampling rate (1000 SPS). Since the wireless data transmission is faster than the data rate, these extra cycles are very important in terms of computation, allowing the microcontroller to properly encapsulate and transmit the messages before a new conversion of sEMG data. Therefore, this communication protocol is important to ensure the 1000 Hz sampling frequency of the sEMG device.

### 3.3. Software Programming and Signal Processing

A firmware was developed and embedded in the microcontroller to encrypt data according to the aforementioned protocol. The data is then transmitted to a personal computer where it is decrypted and presented to the user.

LabVIEW™ is a visual programming environment for developing interactive applications based on virtual instruments (VI) for processing routines. The VI subdivides the application into two parts: block diagram and front panel. The block diagram contains the programming logic, created by manipulating block functions. Likewise, the front panel contains user interface functions such as buttons, indicators, controllers and graphics. Therefore, a LabVIEW™ application was developed for the proposed sEMG device, considering that this environment delivers the opportunity to build user interfaces as well as manage data in binary level. Alternatively, other signal and data processing software (such as Matlab®) could be used as well, as long as they also offer a user interface development environment (which is a feature more optimized and versatile in LabVIEW™ than in Matlab®).

The algorithm developed in LabVIEW™ is responsible for further processing sEMG data. Figure 6 presents a block diagram of the main routines in the VI. First, a serial communication block function is necessary to establish a connection with the Bluetooth module. Once connected, data is read and the decryption process begins.

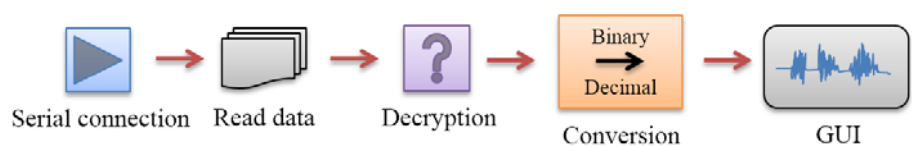

**Figure 6**. Main routine of the VI.

By means of binary operations, the decryption method reassembles a channel by recombining the MSB and LSB of the data window. Afterward, the sEMG data is converted from binary to decimal and displayed in a graphical user interface.

### 3.4. sEMG Device

Subsequently, the developed sEMG device presented in Figure 7 integrates both the hardware (a) and the software (b) components as a way to achieve muscular signal acquisition. In particular, the Bluetooth module and the microcontroller are peripherals that are attached to the board through connectors and can be removed as necessary for debugging or inspection.

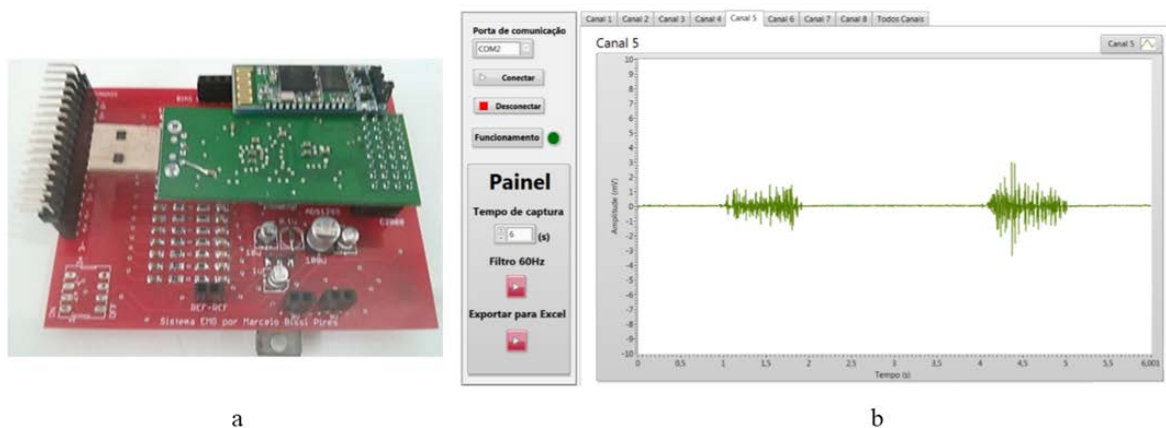

**Figure 7**. sEMG device: (**a**) hardware on the PCB and (**b**) front panel of the GUI.

Additionally, the front panel of the VI plots the eight sEMG channels in real time while granting the following controls: start/stop recordings, switch on/off an online 60 Hz notch filter, save data, and set total acquisition time.

### 3.5. Methodology

After the study of amplitudes and the system resolution, signal conditioning, and communication protocol development, the next step of this work is the experimental methodology with the validation of electronic circuit and the sEMG signal acquisition. The validation process consists the use of known signals (as sinusoidal signal and a sEMG signal from a database) and analyze of its correspondence. The subsequent stage, the sEMG signal acquisition, bases on signal acquisition from arm muscles. These muscles provide easy location and allow the antagonistic movements analyze (as flexion and extension) in a simplified procedure.

Prior to recording an original sEMG signal, the two validation processes were performed to evaluate the relationship between an analog input and the digital output, assessing the accuracy of the device. For the first validation, a function generator was used to create an analog sine wave (frequency of 1 Hz) with an appropriate amplitude range to sweep the full-scale range of the sEMG device (bypassing the hardware filter).

For the second validation, a known sEMG signal from the arm obtained from an online database [29] was converted from digital to analog format with the help of a data acquisition device (NI-DAQ 6211). This analog sEMG signal was then applied to the device without bypassing the hardware filter.

On the other stage of methodology, surface electromyography data was recorded from the left arm of a single subject. In total, 16 electrodes for all eight channels (in differential mode) were used with the addition of a reference electrode attached to the elbow. In this scenario, Figure 8 depicts the electrodes placement sites for each channel (C1-C8) and the reference electrode (Ref). In addition, the channels are spaced 4 cm from each other. Similarly, the distance of the differential electrode inputs is 2 cm.

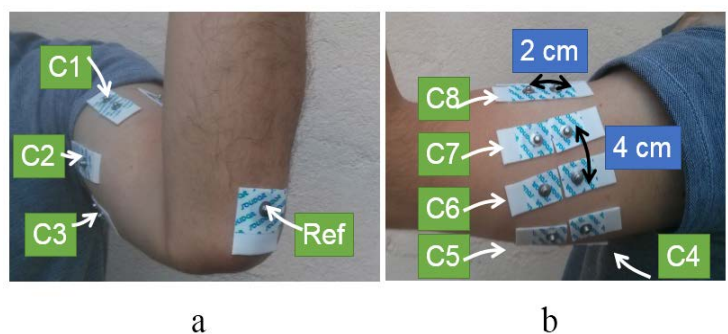

**Figure 8**. Electrodes sites: (**a**) front view and (**b**) side view.

In order to obtain isolated muscle activity (isometric contraction), a specific set of actions was designed as shown in Figure 9. Considering that each step is time limited, the total acquisition time was set to 6 seconds in the front panel of the GUI. Therefore, the subject performed this protocol twice, emphasizing a different muscle on each acquisition: first the triceps and then the biceps. For all the recordings, the 60 Hz notch filter from the GUI was not used in order to assess the existence of this interference on the device.

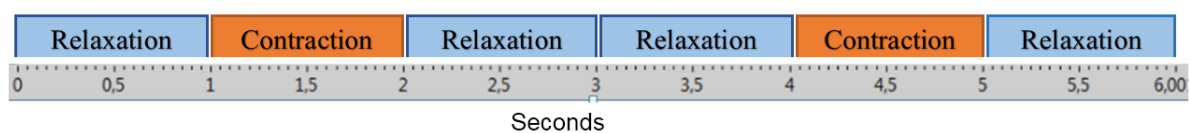

**Figure 9**. Muscle actions.

**4. Results and Discussion**

The results presented and discussed in this section (for both validation and muscular activity recordings) were all obtained with the proposed and developed sEMG device.

*4.1. Validation Process*

The Figure 10a depicts the results of the validation process for the sine wave. The first trace (superior) represents the generated 1 Hz sine wave with a maximized amplitude range of ±17.63 mV, which satisfies by a wide margin the sEMG signals amplitude range [13].

In the second trace (inferior), a full-scale range of ± 17.56 mV can be observed for the validated signal. This amplitude resides in the proposed range of the selected data window (± 17.57 mV), validating the full-scale range of the sEMG device. This implies that the sEMG device (ADS1299 front end) attenuates the signal by a little and, in this particular case, the original-to-reproduced signal ratio of the full-scale range is approximately 99.66% mV (attenuation of 0.030 dB).

Similarly, the superior trace in Figure 10b represents a known raw sEMG signal (originally sampled at 1000 Hz), which was converted to analog and connected to the sEMG device.

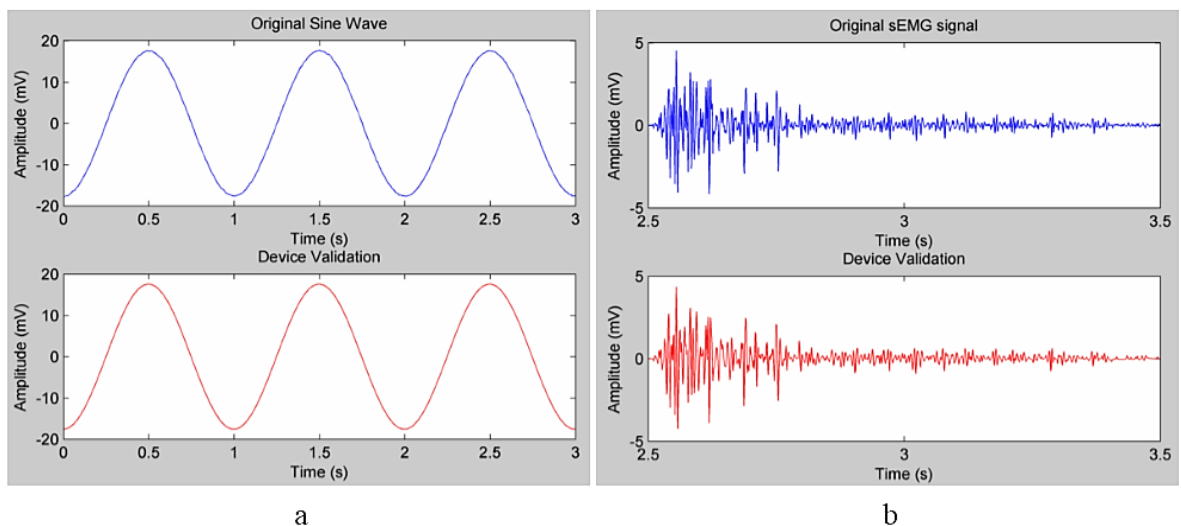

**Figure 10.** Signals of validation process: (**a**) sine wave and (**b**) sEMG signals.

Furthermore, Figure 10b presents in the inferior traces the outputs of the sEMG device. Comparing the two sets of data numerically results in a mean squared error of 0.0879 mV. It is worth mentioning that a main portion of this error is by virtue of comparing a raw signal (the original) with a band-pass filtered one (device validation). Therefore, the reproduction of the sEMG signal showed to be accurate.

*4.2.   Experimental Results (Amplitude)*

All the eight channels successfully recorded muscular activity from the arm on the two acquisitions. However, the electrodes located right above the biceps/triceps delivered the highest potentials. Therefore, once evaluated all the data, it was noted that for the triceps acquisition, channels 5 and 6 exhibited greater muscle activity as presented in Figure 11a. Likewise, for the biceps acquisition, channels 1 and 2 contributed the most for the sEMG signal as depicted in Figure 11b.

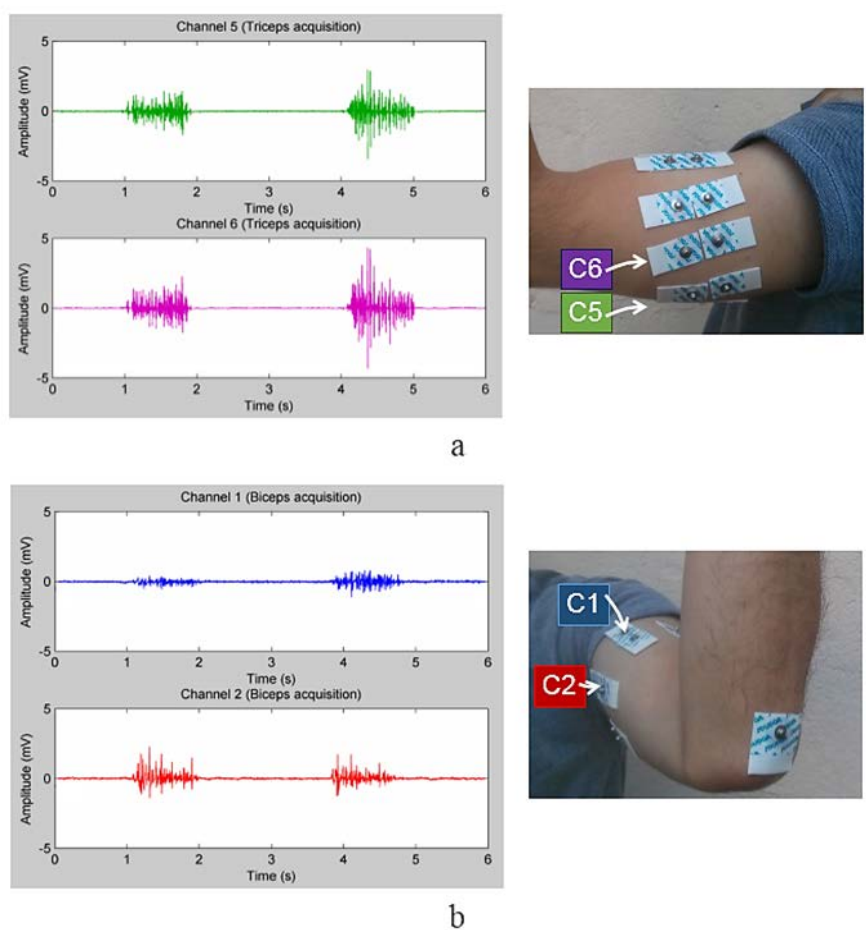

**Figure 11**. sEMG data for (**a**) triceps and (**b**) biceps acquisitions.

For this subject, the triceps contractions generated potentials in the vicinity of ± 5 mV. In contrast, the biceps channels produced lower potentials, around ± 2 mV.

## *4.3. Experimental Results (Frequency)*

On a different note, the frequency spectrum was analyzed based on channel 6 (triceps acquisition) and channel 2 data (biceps acquisition). From Figure 12, the largest amplitudes of the sEMG signal are on frequencies smaller than ~300 Hz for the triceps and smaller than ~220 Hz for the biceps.

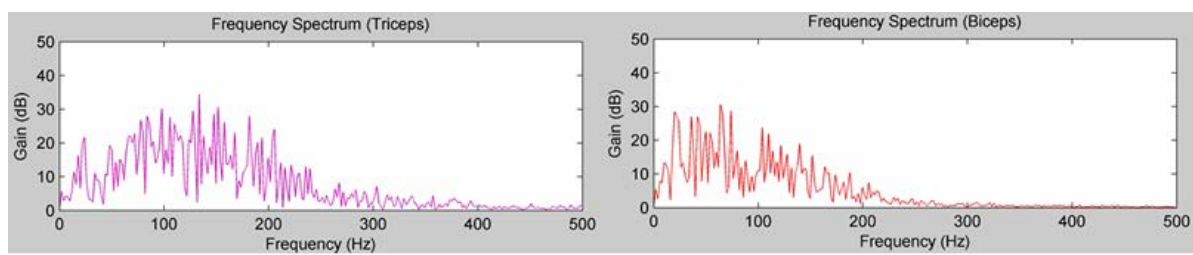

**Figure 12**. Frequency spectrum for the triceps and biceps.

Analyzing the 60 Hz frequency component of the data window corresponding to the absence of muscular activity (rest period) revealed a variety of gains from one channel to another. Channel 4 presented the lowest interference value (0.35 dB) whilst channel 8 presented the highest (7.22 dB).

It is possible that the different susceptibility to noises corresponds to the poor quality of electrodes cables and connectors used (handmade). Therefore, the 60 Hz component from the power line distribution causes little interference on this device, which can be further attenuated by the 60 Hz notch-filter in the GUI.

## *4.4. Comparison of results*

This work has information to increase the circuit reproducibility in virtue of experimental methodology process developed, allowing its replication. It is not found on the works with ADS1299 on literature, which only presents the obtained results and not how the circuit/system is done in all development process. The Table 3 presents a comparison of obtained results with this application with others previously mentioned with using the ADS1299 for sEMG acquisition.

**Table 3.** Comparison among the ADS1299 devices for sEMG with the proposed device.

| A/Y | Purpose | CH | FBW | RES | SR | GUI | BS | μC | WR |
|---|---|---|---|---|---|---|---|---|---|
| [11] 2015 | sEMG | 8 | - | 24 bits | 500 Hz | - | Yes | - | Bluetooth |
| [20] 2015 | sEMG | 4 | 5 Hz-∞ | - | 1 kHz | - | - | MSP430 | Bluetooth |
| [22] 2015 | sEMG | 4 | 5-450 Hz | - | 2 kHz | - | No | - | No |
| [24] 2013 | sEMG | 4 | - | 24 bits | 1 kHz | LabVIEW | Yes | Arduino Uno | WiFi |
| Proposed | sEMG | 8 | 7-338 Hz | 10 bits | 1 kHz | LabVIEW | Yes | TMS320F28069 | Bluetooth |

The most of aforementioned application described on Table 3 acquire the sEMG using 4 channels, whilst the developed device uses 8 channels, increasing the number of signals to analyses. The filters provides the acquisition in the bandwidth of 7-338 Hz, inside of sEMG bandwidth. There are applications without filters, even though it is necessary to attenuate noises. The proposed system has a 10-bit resolution, with each bit equals to 34.33 µV in a range of ±17.57 mV. In the others works are presented 24-bit resolutions, but the reference voltages to perform this calculation are not mentioned in the most papers. A sample rating for this application is inside of recommendable range for sEMG signal, being compatible with the similar works. About the GUI, only one presents this functionality also with LabVIEW. The battery supply is present in generally works as also the specification of microcontroller. At least, similar works use the Bluetooth technology to data communication, guarantee the connectivity since smartphones to computers using wireless communication.

## 5. Conclusion

This work presented a portable, wireless sEMG device using the ADS1299 analog front-end (commonly used in EEG and ECG applications). The unique user interface developed in LabVIEW$^{TM}$ allows users to quickly acquire, save and analyze sEMG data. However, as observed by the validations and the results, this front-end is also capable of capturing muscular activity. Currently, the device is only tested for arm muscles for validation purposes but can easily be applied to other muscle regions as well.

The communication protocol was able to satisfy the sEMG signal characteristics by making both resolution (34.33 µV) and amplitude range (±17.57 mV) compatible. The reduction of the message length to 110 bits provided a faster transmission rate, and as a consequence, the device was able to perform sEMG acquisition at 1000 SPS for all eight channels simultaneously. Therefore, even though the ADS1299 is capable of higher sample rates (from 250 SPS to 16 kSPS), regarding this matter, the protocol presents a restriction to the system, and recordings performed over 1000 SPS result in incorrect readings.

The subsequent development of this work includes some main enhancements. The supply circuit can be improved to make use of only one lithium-ion battery to generate both positive and negative voltages. The attachable peripherals can also be mounted on the PCB instead of being connected externally, increasing the overall quality. All the RC filters can be replaced with active filters to improve the frequency response, the impedance matching and if necessary to provide an amplification to the signal and multiple filtering stages. Moreover, better fabricated electrodes connectors and cables can help to reduce the 60 Hz noise susceptibility.

Finally, the developed sEMG device successfully captured muscular activities, generating enthusiasm and opportunities for further developments. As future work, it presents the possibility of data acquisition for Android mobile devices and thus enable data acquisition during long shifts.

**References**

1. Fall, C. L.; Gagnon-Turcotte, G.; Dube, J.-F.; Gagne, J. S.; Delisle, Y.; Campeau-Lecours, A.; Gosselin, C.; Gosselin, B. Wireless sEMG-Based Body-Machine Interface for Assistive Technology Devices. *IEEE J. Biomed. Health Inform.* **2017**, *21*, 967–977, doi:10.1109/JBHI.2016.2642837.

2. Gao, N.; Zhao, L. A Pedestrian Dead Reckoning system using SEMG based on activities recognition. In *2016 IEEE Chinese Guidance, Navigation and Control Conference (CGNCC)*; 2016; pp. 2361–2365.

3. Hung, C. C.; Shen, T. W.; Liang, C. C.; Wu, W. T. Using surface electromyography (SEMG) to classify low back pain based on lifting capacity evaluation with principal component analysis neural network method. In *2014 36th Annual International Conference of the IEEE Engineering in Medicine and Biology Society*; 2014; pp. 18–21.

4.  Sarmiento, J. F.; Bastos, T. F.; Botti, A. B.; Elias, A.; Frizera, A.; Hubner, M.; Silva, I. V. Characterization and diagnosis of fibromyalgia based on fatigue analysis with sEMG signals. In *2012 ISSNIP Biosignals and Biorobotics Conference: Biosignals and Robotics for Better and Safer Living (BRC)*; 2012; pp. 1–4.

5.  Palmes, P.; Ang, W. T.; Widjaja, F.; Tan, L. C.; Au, W. L. Pattern Mining of Multichannel sEMG for Tremor Classification. *IEEE Trans. Biomed. Eng.* **2010**, *57*, 2795–2805, doi:10.1109/TBME.2010.2076810.

6.  Gutiérrez, S. J.; Cardiel, E.; Hernández, P. R. A muscle fatigue monitor based on the surface electromyography signals and frequency analysis. In *2016 Global Medical Engineering Physics Exchanges/Pan American Health Care Exchanges (GMEPE/PAHCE)*; 2016; pp. 1–6.

7.  Fan, Z.; Zhao, C.; Luo, L.; Huang, S. Study on sEMG-based exercise therapy for upper limb of severe hemiplegic patients. In *2013 35th Annual International Conference of the IEEE Engineering in Medicine and Biology Society (EMBC)*; 2013; pp. 6643–6646.

8.  Tello, R. M. G.; Bastos-Filho, T.; Frizera-Neto, A.; Arjunan, S.; Kumar, D. K. Feature extraction and classification of sEMG signals applied to a virtual hand prosthesis. *Conf. Proc. Annu. Int. Conf. IEEE Eng. Med. Biol. Soc. IEEE Eng. Med. Biol. Soc. Annu. Conf.* **2013**, *2013*, 1911–1914, doi:10.1109/EMBC.2013.6609899.

9.  Kosmidou, V. E.; Hadjileontiadis, L. J. Sign Language Recognition Using Intrinsic-Mode Sample Entropy on sEMG and Accelerometer Data. *IEEE Trans. Biomed. Eng.* **2009**, *56*, 2879–2890, doi:10.1109/TBME.2009.2013200.

10. Ramos, J. L. A. de Controle de Torque de um Exoesqueleto Atuado por Músculos Pneumáticos Artificiais Utilizando Sinais Eletromiográficos. Dissertação (Mestrado em Engenharia Mecânica), Pontifícia Universidade Católica do Rio de Janeiro: Rio de Janeiro, 2013.

11. Jiang, M.; Gia, T. N.; Anzanpour, A.; Rahmani, A. M.; Westerlund, T.; Salanterä, S.; Liljeberg, P.; Tenhunen, H. IoT-based remote facial expression monitoring system with sEMG signal. In *2016 IEEE Sensors Applications Symposium (SAS)*; 2016; pp. 1–6.

12. Fiorucci, E.; Bucci, G.; Cattaneo, R.; Monaco, A. The Measurement of Surface Electromyographic Signal in Rest Position for the Correct Prescription of Eyeglasses. *IEEE Trans. Instrum. Meas.* **2012**, *61*, 419–428, doi:10.1109/TIM.2011.2164838.

13. Hakonen, M.; Piitulainen, H.; Visala, A. Current state of digital signal processing in myoelectric interfaces and related applications. *Biomed. Signal Process. Control* **2015**, *18*, 334–359, doi:10.1016/j.bspc.2015.02.009.

14. Teja, S. S. S.; Embrandiri, S. S.; Chandrachoodan, N.; M, R. R. EOG based virtual keyboard. In *2015 41st Annual Northeast Biomedical Engineering Conference (NEBEC)*; 2015; pp. 1–2.

15. Chen, J.; Li, X.; Mi, X.; Pan, S. A high precision EEG acquisition system based on the CompactPCI platform. In *2014 7th International Conference on Biomedical Engineering and Informatics*; 2014; pp. 511–516.

16. Davies, P. J.; Bohórquez, J. Design of a Portable Wireless EEG System Using a Fully Integrated Analog Front End. In *2013 29th Southern Biomedical Engineering Conference*; 2013; pp. 63–64.

17. Cui, X.; Yang, P. The Front-End Design of Portable EEG Acquisition System Based On the ADS1299. *Int. J. Sci. Res.* **2016**, *5*.

18. Toresano, L. O. H. Z.; Wijaya, S. K.; Prawito; Sudarmaji, A.; Syakura, A.; Badri, C. Data acquisition instrument for EEG based on embedded system. *AIP Conf. Proc.* **2017**, *1817*, 040009, doi:10.1063/1.4976794.

19. Lovelace, J. A.; Witt, T. S.; Beyette, F. R. Bluetooth enabled electroencephalograph (EEG) platform. In *2013 IEEE 56th International Midwest Symposium on Circuits and Systems (MWSCAS)*; 2013; pp. 1172–1175.

20. Wu, J.; Tian, Z.; Sun, L.; Estevez, L.; Jafari, R. Real-time American Sign Language Recognition using wrist-worn motion and surface EMG sensors. In *2015 IEEE 12th International Conference on Wearable and Implantable Body Sensor Networks (BSN)*; 2015; pp. 1–6.

21. Zanetti, R.; Assunção, M. L. M.; Corrêa, M. F. S.; Tierra-Criolo, C. J.; Melges, D. B. Sistema de Aquisição De Sinais Biomédicos Baseado No Front-end ADS1299. In *Proceedings of the XXIV Brazilian Congress on Biomedical Engineering*; Uberlândia, Brazil, 2014; pp. 1649–1652.

22. Guerrero, F. N.; Spinelli, E. Surface EMG Multichannel Measurements Using Active, Dry Branched Electrodes. In *VI Latin American Congress on Biomedical Engineering CLAIB 2014, Paraná, Argentina 29, 30 & 31 October 2014*; IFMBE Proceedings; Springer, Cham, 2015; pp. 1–4 ISBN 978-3-319-13116-0.

23. Bai, O.; Atri, R.; Marquez, J. S.; Fei, D. Y. Characterization of lower limb activity during gait using wearable, multi-channel surface EMG and IMU sensors. In *2017 International Electrical Engineering Congress (iEECON)*; 2017; pp. 1–4.

24. Su, Y.; Routhu, S.; Aydinalp, C.; Moon, K.; Ozturk, Y. Low Power Spinal Motion and Muscle Activity Monitor. In *2015 IEEE Global Communications Conference (GLOBECOM)*; 2015; pp. 1–5.

25. De Luca, C. J.; Gilmore, L. D.; Kuznetsov, M.; Roy, S. H. Filtering the surface EMG signal: Movement artifact and baseline noise contamination. *J. Biomech.* **2010**, *43*, 1573–1579.

26. Texas Instruments. ADS1299-x Low-Noise, 4-, 6-, 8-Channel, 4-Bit, Analog-to-Digital Converter for EEG and Biopotential Measurements Available online: http://www.ti.com/lit/ds/symlink/ads1299.pdf (accessed on May 12, 2017).

27. Guangzhou HC. JYMCU HC06 Product Data Sheet Available online: http://silabs.org.ua/bc4/hc06.pdf (accessed on May 12, 2017).

28. Texas Instruments. TMS320x2806x Piccolo Technical Reference Manual Available online: http://www.ti.com/product/TMS320F28069/datasheet (accessed on May 12, 2017).

29. FeatureFinder. Sample Data Available online: http://www.featurefinder.ca/sample-data/ (accessed on Jul 4, 2017).